\documentclass[10pt,letterpaper]{article}
\usepackage{opex3}
\begin{document}
\title{Theory of unconventional Smith-Purcell radiation in finite-size 
photonic crystals}
\author{Tetsuyuki Ochiai${}^1$ and Kazuo Ohtaka${}^2$}
\address{${}^1$Quantum Dot Research Center, National Institute for Materials
Science (NIMS), Tsukuba 305-0044, Japan\\
${}^2$Center for Frontier Science, Chiba University, Chiba 263-8522, Japan}
\email{OCHIAI.Tetsuyuki@nims.go.jp}


\begin{abstract}
Unusual emission of light, called the unconventional Smith-Purcell
 radiation (uSPR) in this paper,  was demonstrated from   an   electron
 traveling  near a finite photonic crystal (PhC) at  an
 ultra-relativistic velocity.  
This phenomenon is not related to the accepted mechanism  of the
 conventional SPR and  arises because  the evanescent light from the
 electron  has   such a  small    decay constant  in the
 ultra-relativistic regime that it   works  practically as a  plane-wave
 probe entering the  PhC from one end.   
We analyze     the  dependence of the SPR spectrum on the   velocity of
 electron  and  on  the   parity  of   excited  photonic bands and  show, for  PhCs made up   of a  finite number of  cylinders,  that uSPR  probes  the photonic band structure  very faithfully.
\end{abstract}



\section{Introduction}

A traveling charged particle induces  coherent radiation when it passes
near a periodic dielectric structure along the direction of its spatial 
periodicity.  This radiation,  called  Smith-Purcell radiation
(SPR) \cite{Smith:P::92:p1069:1953,Shestopalov-SPR-book}, can be a  novel 
radiation source  with several remarkable properties.  The most important one is the scalability of the output
frequency; the  threshold
frequency below  which  SPR is kinetically impossible   varies in inverse proportion to the   magnitude of the period.  In addition, the SPR is characterized by the presence of   resonances   at a series of    frequencies, which again vary in inverse proportion to the
period. 
 Owing to these 
properties, it has been recognized that SPR can be  a basic  mechanism for a compact free-electron laser \cite{Wachtel::50:p49-56:1979}.

Since its  first observation, SPR has been studied using mainly 
 metallic diffraction gratings of one-dimensional
periodicity \cite{Wortman:L:D:M::24:p1150-1153:1981,Gover:D:E::1:p723-728:1984,Shih:S:M:C::7:p345-350:1990,Doucas:M:O:W:K::69:p1761-1764:1992,Ishi:S:T:H:I:T:M:K:F::51:pR5212-R5215:1995}.
In most theoretical analyses
made so far,  
the gratings have been treated as  perfect conductors  to
simplify the treatment of the periodic light scattering \cite{vandenBerg::63:p1588:1973,Haeberle:R:S:M::49:p3340-3352:1994,Shibata:H:I:O:I:N:O:U:T:M:K:F::57:p1061-1074:1998,Brownell:W:D::57:p1075-1080:1998}.
In these systems,   Wood's anomaly \cite{Wood::48:p928:1935} in the optical 
density of states (ODOS) is responsible for the   enhanced signals of  SPR, 
and thus the relevant frequencies  of the resonances  are determined in a  straightforward manner 
using  simple kinetics. 
  
Recently, 
both theoretical \cite{Pendry:M::50:p5062-5073:1994,deAbajo::61:p5743-5752:2000,Ohtaka:Y::91:p477-483:2001,Yamaguti:I:H:O::66:p195202:2002,deAbajo:B::67:p125108:2003,Ochiai:O::69:p125106:2004,Ochiai:O::69:p125107:2004} and
experimental \cite{Yamamoto:S:Y:S:S:I:O:H:K:M:H:M:Y:O::69:p045601:2004}  SPR results  have been reported 
 for  photonic crystals (PhCs) used in place of metal gratings.   It was found  that PhCs induced  highly  coherent SPR because of their multidimensional periodicity 
in their dielectric functions.
The SPR spectrum  consists of   point-like signals as a function of frequency, which show up each time the evanescent light from the electron excites 
a  photonic band (PhB) mode of high quality factor. 
 SPR from a PhC is versatile, because PhCs  generally have   various  
parameters,  which  can now be reliably designed and  changed.

However, when an electron beam of ultra-relativistic  velocity was used in combination  with   a  PhC, which was   finite   in the direction of the electron trajectory, unexpected phenomena that contradicted  the conventional 
understanding of the SPR were experimentally observed \cite{Horiuchi-unpub}. 
Such phenomena  have not been  observed in the gratings of nearly perfect conductors 
and are expected to be 
absent  even in PhCs  when  an  electron beam of slower velocity is used.   
This SPR,  called   unconventional SPR (uSPR) in this paper, is quite distinct in many ways from the
conventional SPR (cSPR)
 and thus is easily  identified; most importantly,  the   uSPR spectrum
 sweeps  the entire region of frequency-momentum phase space,   in
 contrast to the cSPR  which carries information of the phase space
 only  along the shifted $v$ lines, to be defined later.   In the phase space, the uSPR is   characterized    by 
 peculiar  resonances  arising along  curves,  which are related more or
 less to  the dispersion relations of PhBs, not to the shifted $v$
 lines,  with relatively little intensity variation.  Therefore,   SPR
 in   the ultra-relativistic regime of the beam velocity is potentially
 useful both as a   monochromatic light source  and   as a probe to
 investigate  the PhB structure.   
We should note here that the Cherenkov effect also can be used to probe
the PhB structure with the angle-resolved electron energy loss
spectroscopy \cite{deAbajo:P:Z:R:W:E::91:p143902:2003,deAbajo:R:Z:E::68:p205105:2003,Luo:I:J:J::299:p368-371:2003}.

The conjecture inferred by the experiment, which this paper  seeks to verify,  was that 
the SPR consists of the cSPR and uSPR in the relativistic regime of electron velocity 
and that  the uSPR is expected to  disappear gradually with decreasing velocity, to leave  solely the   cSPR component   for     velocities typically less than  a few hundred keV.

Very recently, Kesar {\it et al.} \cite{Kesar:H:K:T::71:p016501:2005} reported a systematic discrepancy
between the calculated SPR spectrum in a  finite-size grating  
and that of the conventional theory assuming infinite size. Since they
focused on the diffraction grating of a perfect conductor,  their
discrepancy is not directly related to ours, which arises in systems  involving PhCs with a  finite dielectric
 constant.  As for this point, it is noteworthy that a rigorous theory was developed for finite-size grating of infinitely-thin metallic plates \cite{Shestopalov-SPR-book}.

This paper presents  a comprehensive theoretical analysis of the uSPR 
 for the PhCs composed of a finite number of 
cylinders.   We use the multiple-scattering method of theoretical treatment \cite{Ochiai:O::69:p125106:2004}, which  explicitly takes into account the finiteness of the total number of cylinders and  treats exactly the multiple Mie scattering among  them.
We shall  investigate the  
properties of the uSPR and the interplay between u and c SPRs by
 changing various parameters,  such as electron velocity, dielectric
 constant of PhCs, length of PhCs, thickness of  PhCs and  angle of SPR emission. 
Most  of this
paper will  focus on the PhCs of dielectric  cylinders of circular cross section. 
The remarkable 
difference of the present work from  the previous theoretical works involving PhCs lies in the
fact that  we are dealing with the  finiteness of the length of PhC in
 the direction of the electron beam.

This paper is organized as follows.  
In Sec. II, we present  the kinetics for  both  cSPR and
uSPR  and discuss what plays a 
key role in inducing uSPR. 
Section III is devoted to the comparison of the spectrum  of  uSPR with that of cSPR.  
In Sec. IV, we present detailed  analyses of  uSPR  
by changing various parameters.  Finally, we summarize the results in
Sec. V.

\section{Kinetics of conventional and unconventional SPRs}

\begin{figure}
\begin{center}
\includegraphics*[width=0.6\textwidth]{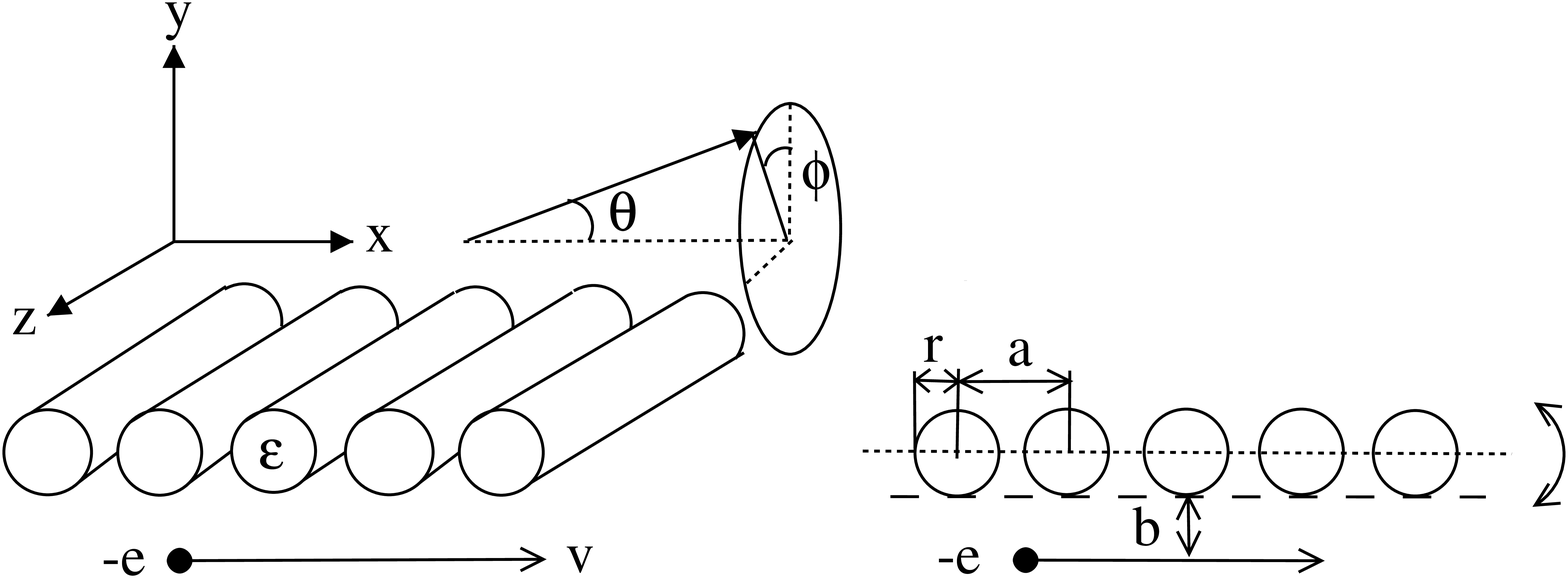}
\end{center}
\caption{\label{geometry}Schematic illustration of the system under
 study. An electron travels with constant velocity $v$ and 
impact parameter $b$ below the
 one-dimensional periodic array of dielectric or metallic cylinders.
 The cylinders  are arrayed periodically with axes in the $z$ 
direction with  radius   $r$,  dielectric constant  $\varepsilon$
and the lattice constant  $a$. The trajectory of the electron
is parallel to the direction of the periodicity. Angles $\theta$ and $\phi$ are the polar angles of the SPR signals.
This PhC has a mirror plane indicated 
by dotted line in the right panel.}
\end{figure}

In the following discussion, we focus on a PhC composed
of infinitely long cylinders.  The system under study is schematically
illustrated  in Fig. \ref{geometry}.
Cylinders are  arrayed  periodically (lattice constant: $a$) in the $x$
 direction with cylinder axes  in the $z$ direction. 
An electron travels near the PhC in a trajectory 
  parallel to   the $x$ axis with  velocity $v$ and 
impact parameter  $b$. 
We obtain the SPR spectrum from this  system  as a sum of the plane-wave signals generated by the scattering of the evanescent light emitted by the electron. The whole process of  multiple  scattering among a finite number of cylinders is dealt with compactly  by  the
multiple-scattering theory of  radiation using  the  vector 
cylindrical waves as a basis of representation \cite{Ochiai:O::69:p125106:2004}.

Let us briefly summarize the kinetics in the  theory of cSPR. In the theory of cSPR, a finite periodic structure   is simulated by  a periodic structure   of infinite length in the $x$ direction.  A
traveling  
electron accompanies the radiation field  
that  is a superposition 
of evanescent waves with respect to frequency $\omega$ and 
wave number $k_z$ in the $z$ direction \cite{Ohtaka:Y::91:p477-483:2001}.  
The wave vector of each evanescent wave is given by 
\begin{eqnarray}
& & {\bf K}^\pm=\left({\omega\over v},\pm\Gamma,k_z \right), \quad 
\Gamma=\sqrt{\left({\omega\over c}\right)^2
               -\left({\omega\over v}\right)^2-k_z^2}, \label{Gamma}
\end{eqnarray} 
where  $\Gamma$ is purely imaginary because $v \le c$.  
The imaginary part $|\Gamma|$ determines  the spatial decay  of  the
evanescent wave incident on the PhC. 
 In what follows,   it is important  to remember the feature of $k_z=0$
  that, in the ultimate  limit $v \to c$, $|\Gamma|$ tends to zero. 
Since $\omega$ and $k_z$ are conserved quantities in the geometry of
Fig. \ref{geometry},
the evanescent waves with different $\omega$ and $k_z$ 
are  independent in the whole scattering process. Thus,  the incident
light of $\omega$ and $k_z$ leaves  the PhC, after being scattered, with
the same $\omega$ and $k_z$.  Therefore,   the SPR signals  observed   in the $xy$ plane may be analyzed by  setting  $k_z=0$ everywhere.       
Since we are now dealing with a perfect periodicity extending from $-\infty$ to $\infty$, we  obtain the  SPR  signal in the form of 
Bragg-scattered waves summed over  the diffraction channels.   The channels are   specified by the  wave vector  ${\bf K}_h^\pm $
defined by   
\begin{eqnarray}
& &{\bf K}_h^\pm=\left({\omega\over v}-h,\pm\Gamma_h,k_z \right),\label{khpm}
\quad \Gamma_h=\sqrt{\left({\omega\over c}\right)^2-\left({\omega\over v}-h\right)^2-k_z^2}. 
\end{eqnarray}
Here, $h=2\pi n/a \; (n: {\rm integer})$ is a reciprocal lattice point 
of the PhC in the $x$ direction. 
Before the scattering by the PhC, $\omega$ and $k_x$, the $x$ component
of the wavevector of light, satisfy the relation   $k_x=\omega/v$. 
The line $\omega=v k_x$,  called   the $v$ line in this paper, lies outside the light cone in the phase space  $(k_x, \omega)$.
After the scattering, the $x$ component of the light of  channel $h$
becomes $k_x=\omega/v-h$. 
  The shifted $v$ line defined by this equation 
is  inside the light cone in a certain frequency
range.    In that frequency range  we can detect the SPR signal in this channel
at a far-field observation point.  The propagating direction of the SPR signal 
of  $\omega$ is given by 
\begin{eqnarray}
{\bf K}_h^\pm =
\frac{\omega}{c}(\cos\theta,\sin\theta\cos\phi,\sin\theta\sin\phi), 
\label{polar-angle}
\end{eqnarray}
in the polar coordinates defined in Fig. \ref{geometry}.

The inside region of the light cone is the leaky region of PhBs, and, accordingly,  the ODOS 
is  nontrivial  there. Actually, 
ODOS has  a sequence of peaks of finite width in the $(k_x,k_z, \omega)$ space. The peak position determines  the  
dispersion relations $\omega=\omega_n(k_x,k_z)$
 of the quasi-guided PhB modes. 
 Imagine temporarily $k_z=0$, for brevity.  The presence of the  modes in the $(k_x, \omega)$ space  significantly affects the SPR spectrum by causing  a sharp resonance
when the dispersion curves of PhBs intersect the shifted $v$ lines. The resonance becomes sharper as the quality
factor of the relevant PhB modes increases \cite{Ohtaka:Y::91:p477-483:2001,Yamaguti:I:H:O::66:p195202:2002}.
 Therefore, the SPR from a  PhC  can have very high quality, which is
 intriguing as a new possibility of PhC.

In the above argument,  
we assumed the conservation of $k_x$, with an Umklapp allowance taken into account. 
In an actual PhC with a finite number $N$ of cylinders, the periodicity or 
the translational invariance of the whole system  is lost at the sample edges. 
One way to take account of the finiteness of $N$ is to treat $k_x$ as  defined only approximately with a  width of the order of  
$\Delta k_x\simeq 2\pi/(Na)$ \cite{Shestopalov-SPR-book}.   In this approach,  
the shifted $v$ lines are considered to have  the finite width,  
and the PhB dispersion relation  will become detectable within this allowance centering on   the shifted $v$ lines of open channels.   In reality, however,  the uSPR signals  appear in the phase space $(k_x, \omega)$  with a much  larger distribution  than this  straightforward $1/N$ blurring \cite{Horiuchi-unpub}.

Let us now consider an ultra-relativistic velocity
$v\simeq c$. 
We continue to confine ourselves  to   the measurement within the $xy$
plane, i.e., $k_z=0$.  In this case,  
  $\Gamma$ is almost zero,  and 
the evanescent wave incident on the PhC may be  regarded practically as a
plane wave with its wavevector directed  in  the x-direction. 
\begin{figure}
\begin{center}
\includegraphics*[width=0.6\textwidth]{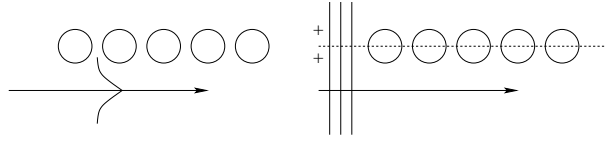}
\end{center}
\caption{\label{unconventional}Schematic illustration of the input evanescent light yielding conventional
 (left panel) and unconventional (right panel) SPRs.  The
 conventional SPR is  produced  when  the evanescent wave has an 
 appreciable decay constant. The incident light   enters 
 the 
 PhC from below. The unconventional case arises when the incident 
 evanescent light has a negligible  decay
 constant  and   is  regarded 
as a plane wave entering  the
 PhC from its  left edge.  In this case,   the evanescent wave is  almost
 symmetric with respect to the mirror plane, inducing the even-parity selection rule in the PhB excitation. } 
\end{figure} 
Therefore, the light-scattering problem is quite similar to that of the light transmission and reflection in the $x$ direction
through the periodic array  of cylinders, as depicted in 
Fig. \ref{unconventional}.  
In this situation, it is obvious that the sample edges play a crucial role.  
In particular, we know that $k_x$ is no longer a 
good quantum number, no matter how large  the total number of the cylinders may be. 
As in an ordinary light-transmission experiment involving PhCs, we expect  the incident wave of  frequency $\omega$ to
excite the PhB modes at the crossing  points between the
line of constant $\omega$ and the dispersion curves
$\omega=\omega_n(k_x,0)$ of PhBs in the $(k_x,\omega)$
plane.  The point is that  the wave vector of the   PhB thus determined is not related to the value of $k_x$ on the $v$ line of the incident light.   Therefore, we should expect  SPR signals over the entire $(k_x, \omega)$ space, not necessarily restricted along the shifted $v$ lines.    The signals expected off the shifted $v$ lines characterize the  uSPR.
Also, we can  expect that the transmitted  SPR  obtained  in the 
side opposite to the  trajectory is almost identical  to
the reflected SPR  of  the  trajectory side.  Finally, similar to the
ordinary  setup of a plane wave  transmitting  through a  two- and three-dimensional PhC \cite{Sakoda-PC-book}, a selection rule must exist 
for  the symmetry of the PhB modes to be excited.  The PhC in our
problem has a symmetry  with respect to the mirror reflection $y \to
-y$ ($y=0$ is the plane bisecting the PhC),  and in the
ultra-relativistic regime the incident light  is of $y$ independent.  Hence,  solely  the PhB modes of even mirror-symmetry are  expected to  participate in the resonant light scattering. The  even-parity  selection rule will thus characterize  uSPR spectra.

According to this  scenario, 
a PhB mode  manifests itself  in the uSPR.  Therefore, 
the uSPR will have    a rather broad band  as a function of
frequency. This is in contrast to the cSPR, in which 
 excited PhB modes  give rise to sharp resonance peaks
 only along   the shifted $v$ lines.
In addition, since $v\simeq c$, the shifted $v$ lines  coincide with
 the threshold lines  for the  opening of a new  Bragg diffraction
 channel.  The channel opening   often leaves  a singular trace due to  Wood's anomaly  
in the line shape of wave scattering. Thus,  on our  shifted $v$ lines, 
Wood's
anomaly will occur,  together with the resonance peaks of the cSPR  associated with the  PhB excitation. 
In this way, the spectra of the c and u SPRs reveal 
a  quite rich structure when the  electron is  ultra-relativistic.

So far, we have concentrate ourselves on the electron traveling parallel
to the $x$ direction, that is, the direction of the periodicity of the PhC 
under consideration. 
If the velocity vector ${\bf v}$ of the electron is given by 
\begin{eqnarray}
{\bf v}=(v_x,0,v_z)=v(\cos\alpha,0,\sin\alpha),
\end{eqnarray} 
the kinetics of SPR changes accordingly.  In particular, the
dominant component of wave number $k_z$ depends on frequency and is given by 
$k_z=v_z \omega/v^2$.  
Within the conventional theory    
the SPR acquires a significant enhancement when the following three 
conditions are fulfilled:   
\begin{eqnarray}
\omega=v_x(k_x-h)+v_zk_z,\; k_z={v_z\over v^2}\omega,\;
\omega=\omega_n(k_x,k_z), 
\end{eqnarray}
where the first equation defines the shifted $v$ line at nonzero $v_z$. 
As in the case of $v_z=0\;(\alpha=0^\circ)$, this scenario of the SPR is insufficient
 for an ultra-relativistic electron. 
In this case the evanescent wave accompanied by the 
electron can be
effectively treated as a plane wave propagating parallel to ${\bf v}$. 
This highlights the role of the sample edge of the   
finite-size PhC, namely,  the broken translational invariance,  and the photonic band modes on the entire plane of  
$k_z=(v_z/v^2)\omega$  are  
excited, not necessary restricted on the shifted $v$ lines. 
In the limiting case of vanishing $v_x \; (\alpha=90^\circ)$, 
the electron travels parallel
to the cylindrical axis.  There, no propagating radiation is
generated from the PhC, as far as the cylinders have infinite length
in the $z$ direction.   
This is due to the perfect translational invariance along the axis.  
In actual PhC, however, this translational invariance is broken,
yielding a sort of the diffraction radiation.  
Thus, as $\alpha$ varies from $0^\circ$ to $90^\circ$, 
the conventional theory of the SPR, which assumes the translational
invariance both in the $x$ and $z$ direction,  predicts a gradual
disappearance of the SPR. On the other hand, the uSPR gives a novel
radiation irrespective of $\alpha$, in which the broken translational
invariance in the $x$ direction is highlighted at small $\alpha$, and 
that in the $z$ direction is highlighted around $\alpha=90^\circ$.

To summarize, the two SPR spectra, c and u SPRs, coexist  
in the ultra-relativistic regime, 
when  a PhC sample is used.   To analyze the experimental signals based on the knowledge on the band structure of photons, the length of the PhC or the total number of cylinders must be finite but large enough.   For a finite system  to have a band structure  comparable 
to that obtained  for an infinite system, 
the periods of,  $N \ge 8$  will be 
enough according to our experience.  In contrast to the value of $N$ in the $x$ direction,  however, 
we are considering a system having   small size 
in the $y$ direction,  such as a PhC made of a monolayer or  stacked layers of several monolayers. 
  The finite size in the $y$ direction needs to be   
considered  explicitly to obtain   the band structure of our PhCs.

\section{Typical example of conventional and unconventional SPRs}

\begin{figure}
\begin{center}
\includegraphics*[width=0.6\textwidth]{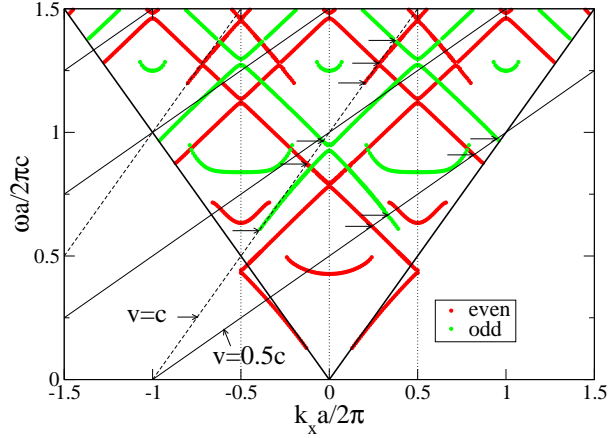}
\end{center}
\caption{\label{teflon_band}  PhB structure of the monolayer of
 low-index  cylinders  in contact ($\varepsilon=2.05$). 
The modes of  TE polarization of  $k_z=0$ are plotted 
 as a function of $k_x$.  
The PhB modes are classified
 according to the parity with respect to the mirror plane bisecting the
 monolayer.
The light line $\omega=\pm ck_x$ is indicated by thick solid lines, the shifted $v$ lines of $v=0.5c$ by thin solid lines, and those
 of $v=0.99999c$  by  dashed lines.   The horizontal  arrows (six for $v=0.99999c$ and four for $v=0.5c$) are drawn  at the intersections between the shifted $v$ lines  and the PhB dispersion curves.  They  correspond to those of Figs. \ref{conv_teflon} and \ref{teflon_v05_Ninf}. 
} 
\end{figure}

Before making a detailed study of  the uSPR properties, we  briefly 
compare the cSPR spectrum with the uSPR spectrum,  using the  numerical results  for a test  system. 
We adopt  a  PhC of a  monolayer of periodic array of  low-index
  cylinders (dielectric constant $\varepsilon=2.05$). 
  For  a radius-to-periodicity ratio $r/a=0.5$, 
 Fig. \ref{teflon_band} depicts the  band structure of the monolayer of an infinite number of cylinders.
The band structure inside the light cone was obtained by plotting the peak
 frequencies of the ODOS, which were   calculated   as a function of 
$k_x$ and $k_z$ \cite{Ohtaka:I:Y::70:p035109:2004}. 
The band structure outside the light cone (that of the true-guided
modes with infinite lifetime) 
was obtained from  the position of the  poles of the  S matrix, which are found  
 on the real axis of  the complex $\omega$ plane.
In Fig. \ref{teflon_band}, $k_z=0$ is assumed, so that the PhB  
modes are decomposed  into purely transverse-electric (TE) and 
transverse-magnetic (TM)  
modes. Only  the band structure of the TE modes
 is presented, because the incident evanescent wave is  
TE-polarized at $k_z=0$.
The PhB modes are further classified by  parity with
respect to the mirror plane $y=0$.  In Fig. \ref{teflon_band} 
the even (odd) parity modes are indicated by red (green) circles. 
We should note that 
 PhB modes having  their dispersion curves disconnected  in Fig. \ref{teflon_band} 
 are the ones  obtained from the  ODOS peaks, which are often too broad to identify   the peak position.

\begin{figure}
\begin{center}
\includegraphics*[width=0.6\textwidth]{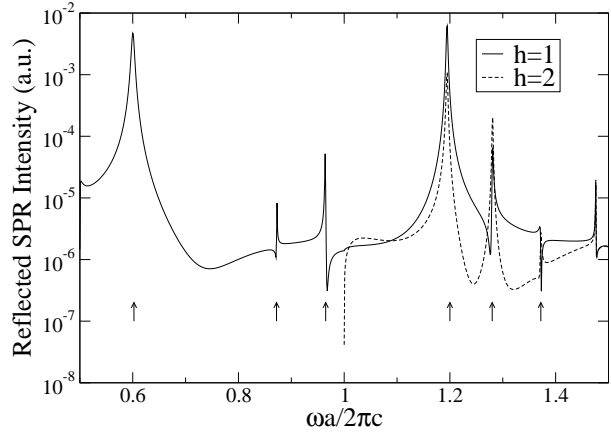}
\end{center}
\caption{\label{conv_teflon} Reflected intensity of cSPR  on the shifted $v$ lines of $v=0.99999c$
 for $\varepsilon=2.05$ and $b=3.33a$.
 Perfect periodicity ranging from $x=-\infty$ to $x=\infty$  of the monolayer cylinders   is assumed.
 The arrows  are drawn  at  the peak positions  and agree with those of
 Fig. \ref{teflon_band}, which were  assigned  to the crossing points between the shifted $v$ line and the band structure.} 
\end{figure}

Let us first consider  the cSPR spectra obtained by the conventional
theory based on  the assumption of    perfect periodicity in the $x$ direction.  We  used the  parameters $v=0.99999c$, $\phi=180^\circ$ and $b=3.33a$  and assumed that 
the radiation was observed in the $xy$ plane $(k_z=0)$, as actually encountered   in   the   millimeter-wave SPR experiments carried out recently \cite{Yamamoto:S:Y:S:S:I:O:H:K:M:H:M:Y:O::69:p045601:2004,Horiuchi-unpub}. 
Since the periodicity is perfect, the SPR spectrum  appears  strictly  on the
shifted $v$ lines, which are almost parallel to the light line
$\omega=ck_x$. 
Figure \ref{conv_teflon} presents the reflected cSPR spectra along the shifted $v$ line of   
$h=1$ and 2 (in units of $2\pi/a$). 
The peaks of the cSPR spectrum arise  
 at the frequency where the shifted $v$ lines $k_x=\omega/v-h$ 
intersect the PhB structure  given  in
Fig. \ref{teflon_band}. Several arrows are drawn  at the peak positions  in
 Fig. \ref{conv_teflon} and, to identify   each of the peaks, horizontal arrows are added in Fig. \ref{teflon_band}  at the corresponding positions in  phase space. 
Comparing these two figures, we see that  the  peak lowest in frequency  arises from  the excitation of an odd-parity PhB mode.  Thus, the even selection-rule for the parity of the excited  PhB modes does not hold for  cSPR, though it somewhat   affects  their spectral shapes.

\begin{figure}
\begin{center}
\includegraphics*[width=0.6\textwidth]{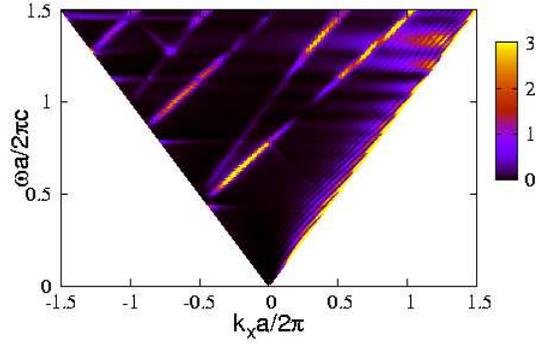}
\end{center}
\caption{\label{unconv_teflon} 
 Reflected SPR intensity map from a finite ($N=21$) monolayer of the 
contact  cylinders with a  low index dielectric constant. The signals appearing off the shifted $v$ lines
 characterize the uSPR.  Except for $N$, the same parameters as in 
Fig. \ref{conv_teflon} were used.  
 } 
\end{figure}

Now we turn to the uSPR spectrum.   The SPR spectrum, with the finiteness of  $N$  considered  explicitly,
 is obtained over the entire $(k_x, \omega)$ space by summing all the amplitudes of the multiply scattered light from the $N$ cylinders \cite{Ochiai:O::69:p125106:2004}.  The numerical result for $N=21$ is given in 
Fig. \ref{unconv_teflon},
for the same parameters of $r/a$ and $\varepsilon$ as used 
 in Fig. \ref{conv_teflon}. 
The angle $\theta$- and frequency $\omega$-resolved reflected SPR 
intensity is mapped onto the 
$(k_x,\omega)$ plane through  the relation $k_x=(\omega/c)\cos\theta$. 
To be precise, $|f^M(\theta)|^2$ with $-\pi \le \theta \le 0$ defined in 
Eq. (33) of Ref.  \cite{Ochiai:O::69:p125106:2004} was plotted by using
the above relation.

\begin{figure}
\begin{center}
\includegraphics*[width=0.6\textwidth]{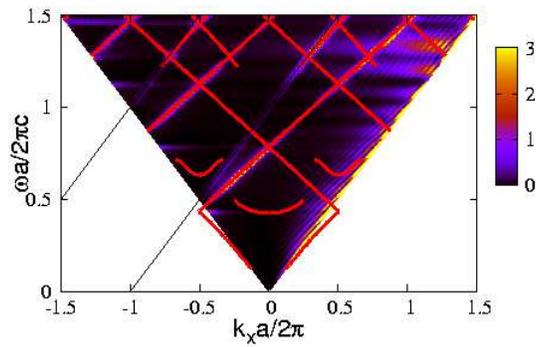}
\end{center}
\caption{\label{unconv_teflon_PB} 
 Figure \ref{unconv_teflon} of the intensity  overlaid with 
the PhB structure of the even parity (indicated by red circles). 
The shifted $v$ lines are indicated by solid lines.  } 
\end{figure}

We observe  that the peaks of the SPR intensity are found along\\
 (A) the shifted $v$ lines, \\
(B) the curves whose slopes are positive and less than 1, \\
(C) the curves whose slopes are negative, \\
(D) the forward light-line ($\omega=ck_x$), and \\
(E) the flat lines terminated  on the backward light-line ($\omega=-ck_x$). \\
For comparison,  we superimpose the even-parity PhB structure of TE modes on the intensity map  and present the result  in 
Fig. \ref{unconv_teflon_PB}.
From Fig. \ref{unconv_teflon},  we find that  type (A)   peaks  are
rather broad along the shifted $v$ lines as a function of frequency, as
compared to the cSPR spectrum shown in Fig. \ref{conv_teflon}. 
A good agreement  between    the PhB dispersion curves and  the high intensity positions of the uSPR
demonstrates  that  the peaks of  type (B) 
are attributed to the quasi-guided PhBs of  even parity.  
Also, no evidence of the excitation of the odd-parity modes is found in 
Figs. \ref{unconv_teflon} and \ref{unconv_teflon_PB}. 
Thus, the even-parity  selection rule of uSPR,   predicted in the previous section,  indeed  holds.  
The peaks of type (C)  are found, for instance, around 
$(k_x a/2\pi,\omega a/2\pi c)=(-0.75,1.25)$, whose origin  is also 
 attributed  to excitation of the even-parity PhB modes. The intensity of these peaks
is, however,  
small,  as compared to that caused by the PhB modes  with  positive slope.  
This seems reasonable,  considering  that the PhBs of  the negative
 group velocity will have suppressed  
 excitation probability  at the left edge of the PhC. 
The peaks of type (D) are inevitable in finite systems; 
the incident wave induces the
quasi-guided waves  in the monolayer, which 
 exit from the right edge, causing 
 a forward-oriented diffraction there.   As a consequence, 
broad peaks of the SPR near the forward light-line emerge. 
Note that the intensity of the type  (D) signal  oscillates as a function of
frequency. This is a  Fabry-Perot oscillation of the signal intensity  with 
  period  decreasing  with increasing $N$. 
Finally, the signals of type (E) are caused by the presence of  pseudo gaps 
in  the PhB structure.  
To see this, we have only to note  in Fig. \ref{unconv_teflon_PB} that  the flat streaks of   type (E) signals appear at  the  gap positions.
When the frequency of the incident wave  lies  in a pseudo gap of
the monolayer, the incident wave cannot penetrate deep inside the PhC and is scattered out of the PhC as a type (E) signal, 
 with a certain angular distribution 
centering on  the backward direction. 
Note that the intense streaks are not related to  the band gaps of odd parity.

The above features of  the uSPR in finite monolayers will remain
unchanged  even for a  semi-infinite monolayer, which  is made  of cylinders of $N=\infty$ but  bounded at one end, because what matters in the above discussion   is the presence of the left  edge of PhC as an  entrance surface of a  wave propagating in the $x$ direction.

Next, we consider  SPR spectrum obtained from  a slower electron in a non-relativistic regime.  The calculation  is made for   $v=0.5c$, which is a
typical value for the electron velocity used  in scanning electron
microscopes. The parameters except $v$ and $b$ are the same.
As above, we compare  two spectra  of   $N=\infty$ and $N=21$. 

\begin{figure}
\begin{center}
\includegraphics*[width=0.6\textwidth]{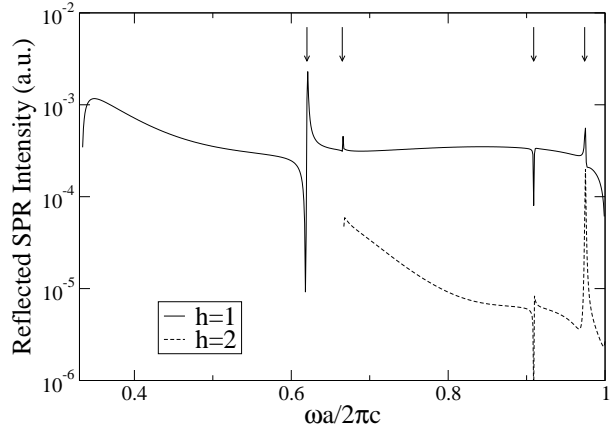}
\end{center}
\caption{\label{teflon_v05_Ninf} 
The reflected cSPR intensity  from the infinite monolayer $(N=\infty)$ of the 
 low-index contact cylinders. The  parameters $v=0.5c$ and $b=0.2a$ were assumed for
the electron beam. Four arrows are assigned  to the peak positions and
 coincide with those of Fig. \ref{teflon_band}. } 
\end{figure}

 First, the reflected SPR spectrum  for  $N=\infty$ is given 
in Fig. \ref{teflon_v05_Ninf}. 
The spectra reveal a marked resonance at $\omega a/2\pi c=0.621$.
The line shape of the resonance is asymmetric as a function of frequency.  
As indicated by arrows, each  agreeing    precisely with those  given to
the shifted $v$ line of Fig. \ref{teflon_band},  
the cSPR peaks all appear exactly at  the
intersection points  of  the shifted $v$ line of $v=0.5c$ with the PhB
dispersion curves.

\begin{figure}
\begin{center}
\includegraphics*[width=0.6\textwidth]{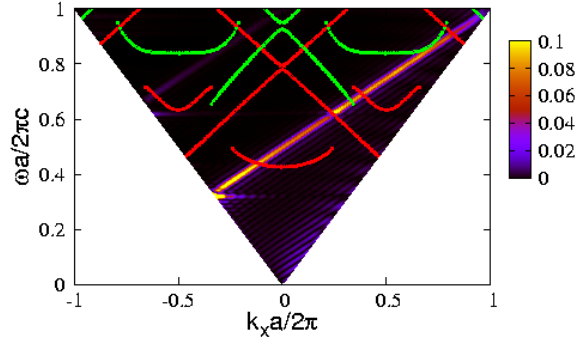}
\end{center}
\caption{\label{teflon_v05_N21} 
 Reflected SPR spectrum  from a finite ($N=21$) monolayer of 
 low-index cylinders in contact.   The intensity profile is overlaid
 with  the corresponding PhB 
 structure of $N=\infty$.  The PhB modes with  even (odd)
 parity are indicated by red (green) circles. 
 The  parameters $v=0.5c$ and $b=0.2a$ were assumed for
the electron beam.} 
\end{figure}

The  reflected SPR spectrum for $N=21$ is given  in Fig. \ref{teflon_v05_N21}, with the superposition of the PhB structure (of $N=\infty$).  
We see at once  that  high  intensity  SPR appears  only  on the shifted
$v$ lines, although very  weak  structures reminiscent of the  finiteness  of our  PhC are still seen off the shifted $v$ line. This is in clear contrast to  the ultra-relativistic
spectra, where marked signals of uSPR existed  definitely off  the
shifted $v$ lines. 
The  signals on  the shifted $v$ line of $h=1$ have  a resonance peak  at 
$(k_xa/2\pi,\omega a/2\pi c)=(0.24,0.62)$. 
This frequency  is almost identical to   that of the  resonance  obtained for $N=\infty$ shown in Fig. \ref{teflon_v05_Ninf}. Also, we can perceive the asymmetry of the line shape  along the shifted $v$ line, as in the
cSPR spectrum of  $N=\infty$.
Therefore, we may conclude that, for   slower velocities such as $v=0.5c$,  the SPR of the finite PhC
can be  understood sufficiently well using the  theory of cSPR,  based on the assumption  $N=\infty$.  The uSPR signals  are suppressed  as follows. The light of $v=0.5c$ is literally  evanescent with an  appreciable   decay constant $|\Gamma|$, so that, while passing through the PhC in the $+y$ direction,  the incident light  decays much and sees  only  the  surface region of cylinders.  Accordingly, the picture of a plane wave with wavevector in the $x$ direction no longer holds and the conventional theory of SPR covers all the features.

\section{Properties of the unconventional SPR}

This section  presents the properties of  
uSPR in detail by changing various  parameters. 
As explained in the previous sections, the broken translational
invariance due to finite number of cylinders ($N$) is crucial in the uSPR. 
Taking account that the uSPR must vanish in the system of the perfect 
translational invariance, it is interesting to investigate the
$N$-dependence of the uSPR in detail. 
The number of stacking layers ($N_l$) is also an important factor because  ODOS
and thus the PhB structure depends crucially on  $N_l$.  
Dielectric constant $\varepsilon$ and radius $r$ of the cylinders are other factors that  
significantly influence the PhB structure.  However, the effects of
changing $r$ are covered, to some extent, by  those  of  $\varepsilon$. 
The impact parameter $b$ is not essential, as seen in the following 
expression for the total emission power $W$ of SPR, whose 
 $b$ dependence is collected into a  simple scaling law \cite{Ochiai:O::69:p125106:2004} 
\begin{eqnarray}
& & W=\int{{\rm d}\omega {\rm d}k_z\over \pi^2} P_{\rm em}(\omega,k_z), \\ 
& & P_{\rm em}(\omega,k_z)|_{b}=e^{-2|\Gamma|(b-b_0)}P_{\rm em}(\omega,k_z)|_{b_0},
\end{eqnarray}
where $P_{\rm em}(\omega,k_z)|_{b}$ is the $\omega$- and $k_z$-resolved emission power for  an impact parameter $b$ and $b_0$ is a reference impact parameter chosen arbitrarily. 
Therefore,  uSPR  and  cSPR  change  in a straightforward way as $b$ varies, with the underlying physics unaltered. 
In the following
subsections, therefore,  five parameters,  $v, \varepsilon, N, N_l$, and $\phi$, 
are varied in this order to 
 see how each  affects the spectrum.

\subsection{Velocity}

The velocity of the electron beam is a key parameter in the
uSPR. Indeed, as  seen in the previous section,  the SPR  at $v=0.5c$ is understood using the theory of cSPR, while at $v=0.99999c$ the uSPR also plays an 
important role. We shall examine  how the conventional picture   fails 
with varying  electron velocity.
An obvious but nonessential  $v$-dependence  is an increase  of the SPR intensity due to  the $v$ dependence of the   decay-constant $|\Gamma|$;  if    impact parameter
$b$ is  fixed,   the overall SPR spectrum   behaves  as $\exp(-2|\Gamma|b)$. To eliminate  this trivial   $v$-dependence,  we have set 
\begin{eqnarray}
b=0.01\beta\gamma a, \quad \mbox{with} \quad \beta={v\over c}, 
\quad \gamma={1\over \sqrt{1-\beta^2}},
\end{eqnarray}
considering $|\Gamma| \propto 1/(\beta\gamma)$ for  $k_z=0$.

\begin{figure*}
\begin{center}
\includegraphics*[width=0.9\textwidth]{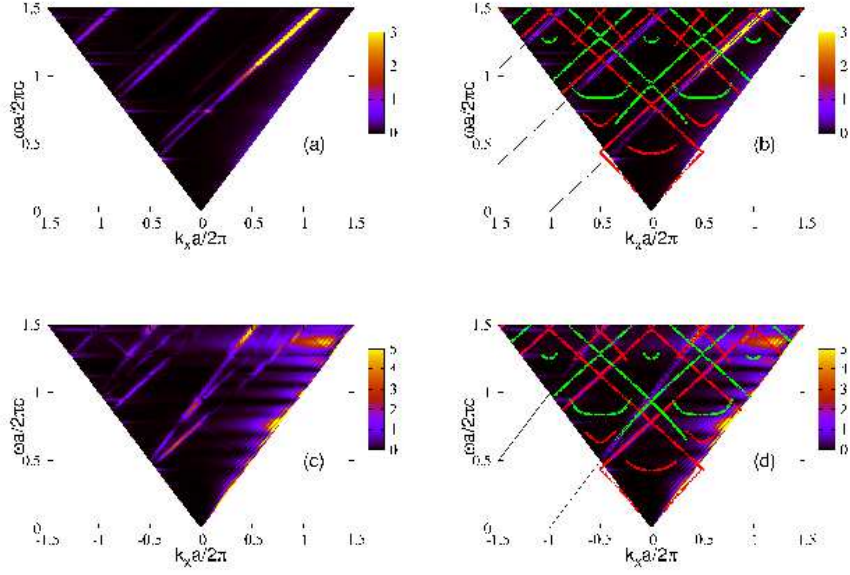}
\end{center}         
\caption{\label{v-change} 
 Dependence of the reflected SPR spectra on electron velocity.  The result for a finite   monolayer ($N=21$)  of  contact  cylinders is shown for $\varepsilon=2.05$.  Panels (a) and (b) are the results for   $v=0.7c$,  and panels (c) and (d) are those for 
 $v=0.99c$.  Panels (b) and (d) are reproductions of  (a) and (c), respectively,  overlaid with  the PhB structure of $N=\infty$.  
The PhB modes with  even (odd) parity are indicated 
by red (green) circles.  See text for
 the  impact parameter $b$ used  in the calculation.} 
\end{figure*}

The reflected  intensity maps for the  monolayer of $N=21$ 
are shown in Figs. \ref{v-change} (a) and (b) for $v=0.7c\quad(\gamma=1.4)$ 
and in (c) and (d) for $0.99c\quad (\gamma=7.09)$, 
along with the PhB structure. 
Panels (a) and (c) show   only  the SPR intensity,  while
 they are  superposed  by 
the PhB structure in  panels (b) and (d).

At $v=0.7c$,  there is a marked bright line  along the shifted
$v$ line of $h=1$. Along  the line, the intensity contrast of the SPR is quite
strong at low frequencies. In particular 
 a point-like resonance is seen  at $\omega a/2\pi c\simeq 0.745$. 
As panel (b) shows,  this resonance arises just  at a crossing between  the  dispersion
curve of an even-parity PhB and the shifted $v$ line of $h=1$. 
Therefore, this  is a  type (A) signal according to  the classification of the last  section. 

 We
can see  the flat streaks  of strong  intensity just at this
frequency,. We note that the signal  becomes stronger as  $k_x$
approaches  the backward light line $\omega=-ck_x$.  
This feature is common to  all the 
horizontal streaks  appearing  at the frequencies of the   pseudo gaps.   These are signals of type (E) of the uSPR. 
We should note that the PhB
mode, which crosses  the shifted $v$ lines,   has a negative group velocity, and the excited  PhB mode  propagates  in the $-x$ direction.   A backward-oriented 
diffraction taking place at the left edge of the PhC explains the tendency towards  the line $\omega=-ck_x$.
  Analogous  flat lines exist, for instance,  at 
$\omega a/2\pi c\simeq 1.09$.

In addition,  we  see clearly   a  high-intensity spectrum  appearing almost  
parallel to the shifted $v$ lines. The curves are in fact coincident with
the dispersion curves of quasi-guided PhB  of the even parity. 
Therefore, they are type (B) signals. Note that  the odd-parity PhB dispersion curves are also visible, with reduced strength as compared to  the even-parity PhBs.     Altogether,  at $v=0.7c$,  cSPR
coexists  with  uSPR and odd-parity PhBs are seen in the uSPR spectrum,
with weaker intensity than even-parity PhBs, however.  Combining this
result with what we have seen in Sec. III  for  $v=0.5c$ and
$v=0.99999c$, 
we may conclude  that, as $v$ increases from $v=0.5c$, the uSPR becomes visible
and the even-parity selection rule of uSPR is less stringent at
non-ultra-relativistic velocities.

The result for $v=0.99c$ indeed confirms this conclusion. 
At $v=0.99c$, several bright curves arise in Figs. \ref{v-change} (c) and (d)  with
little intensity contrast  along  the PhB dispersion curves.  
This is the type (B) signal.  We can observe  odd-parity excitation of 
weak intensity.  Therefore, although
the even-parity  selection rule  is indeed   dominant, it is somewhat
relaxed for $v=0.99c$.  On the shifted $v$ line, there are  signals of
cSPR, as theory predicted for type (A)  features in Sec. III.

The breakdown  of the even-parity selection rule at these velocities 
 is explained  as follows.    With a decrease of  $v$, $|\Gamma|$ increases to make   the incident evanescent light decay more  quickly when  passing the monolayer.   This  increases  the  asymmetry of the evanescent wave with
respect to the mirror plane and  makes
 the even-parity selection rule less effective.   
The degree of the symmetry of the input wave may be given   by  the  factor 
$\exp(-|\Gamma| 2r)$, called  here the symmetry factor,
 which measures  the decay of  the evanescent wave while traversing the PhC in  the $+y$ direction. 
 If this factor  is unity, the evanescent light seen by the PhC  is mirror-symmetric.
At $v=0.99c$, the symmetry factor  is   0.408 at $\omega a/2\pi c=1$ and 
  too small to guarantee strictly the even-parity selection rule.  
Therefore,  odd-parity PhBs are allowed somewhat as uSPR signals.

The results for the other $v$ are briefly summarized without giving numerical results. 
At  $v=0.9c$, when the symmetry factor is 0.047,  cSPR and  uSPR coexist and odd-parity  PhBs are seen in the latter. 
At $v=0.999c$, the asymmetry factor increases to  0.755.  The intensity map gradually tends  to the case of
$v=0.99999c$ with  the symmetry factor  0.972; signals along  the  
odd-parity PhBs disappear, 
 leaving behind only the even-parity signals as type (B) signals.   
The horizontal bright streaks  appear   solely  in the regions of  
 the pseudo-gaps of even-parity bands.

Finally, we should comment on the case of non-zero $v_z$. 
The critical velocity of the electron, above which the uSPR begins to
emerge does not change so much by non-zero $v_z$.  An important point is that 
at ultra-relativistic velocities the evanescent wave can be effectively 
regarded as a plane wave.  
This is not controlled by $v_z$, but is controlled by $v$, the magnitude
of the velocity vector.  However, other features of the uSPR changes 
as discussed in the previous section.

In conclusion, in the frequency region $\omega a/2\pi c \sim 1$, the
uSPR is  conspicuous   when $v$ exceeds 0.7c, and  the even-parity
selection rule  holds  progressively  better as $v$ approaches $c$ from $0.9c$.

\subsection{Dielectric constant}

For a PhC with   $r$ and $N$ kept fixed at $r=0.5a$ and $N=21$, let us
examine  how the SPR spectrum varies as the dielectric constant
$\varepsilon$ of the cylinders changes in the monolayer.   
We select three values of $\varepsilon$,  
$\varepsilon=4.41, \quad 1-(\omega_p/\omega)^2$ and 
$-\infty$.  The first case corresponds  to  the dielectric constant of fused quartz with $\varepsilon$ nearly twice as large as that used above, the second is the dielectric constant of  a Drude
 metal with $\omega_p$  the plasma frequency, 
and the third is the dielectric constant  of  a perfect conductor. 
To avoid the poor convergence of the cylindrical-wave expansion for   the  metallic cylinders in contact, we created  a narrow opening between the   cylinders by setting   $r=0.45a$ in the Drude  case.  We  assumed   
$\omega_p a/2\pi c=1$, i.e., the plasma wavelength 
equals  the lattice constant.    Calculation is made  
 for the monolayer system using  $v=0.99999c$ and  $b=3.33a$, as before.

\begin{figure*}
\begin{center}
\includegraphics*[width=0.9\textwidth]{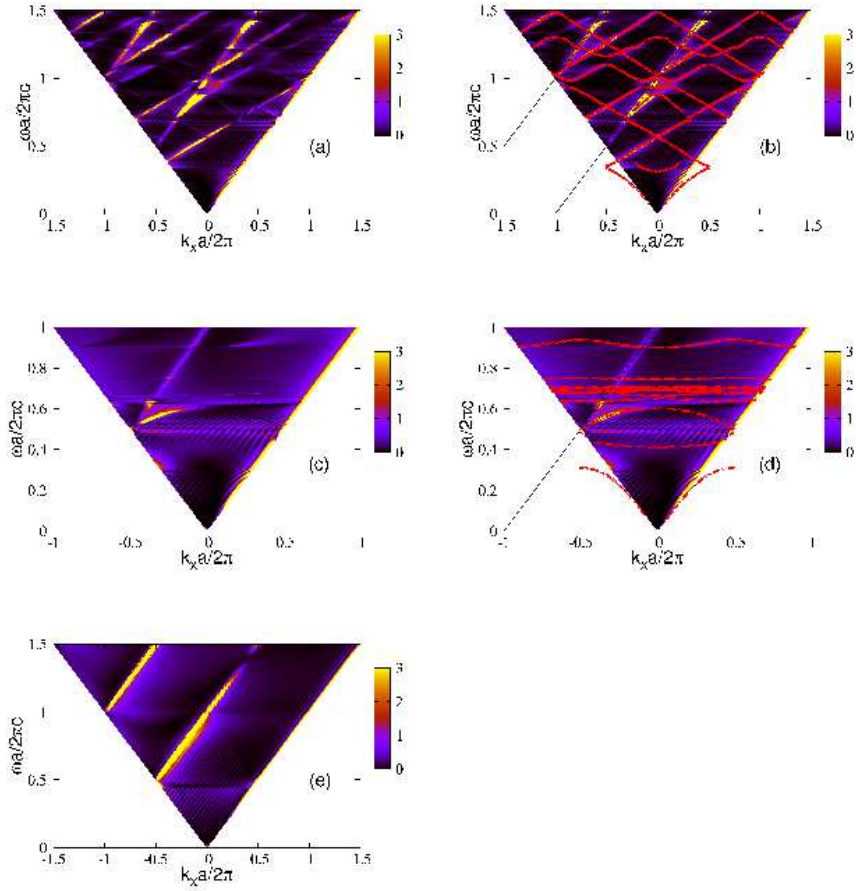}
\end{center}         
\caption{\label{spr_eps} 
 Reflected SPRs from the  monolayers of contact cylinders of various dielectric constants.
   Panel  (a)  shows the result of   dielectric cylinders of $\varepsilon=4.41$ and $r=0.5a$, panel 
(c) is the result of  metallic
cylinders of Drude dielectric constant $\varepsilon
=1-\omega_p^2/\omega^2$ with $\omega_pa/2\pi c=1$, 
and panel (e) shows  the result of 
  cylinders of perfect metal, i.e., $r=0.5a$ and 
$\varepsilon=-\infty$.
 Panels (a) and (c) are reproduced in panels   (b) and (d),
 respectively, with the corresponding PhB superposed. The same parameters as 
in Fig. \ref{conv_teflon} were used for the electron beam.  
} 
\end{figure*}

The reflected SPR intensity maps 
 are shown in Fig. \ref{spr_eps}, together with the corresponding PhB structure obtained for the $N=\infty$ system.
Panels  (a) and (b) depict the result of    dielectric cylinders, 
 panels (c) and (d) treat    the Drude
 cylinders,  and   panel  (e) presents the spectrum  of   the cylinders  of 
 a perfect conductor. Panels  (b) and (d) also involve  the band structures of the monolayer. 
Considering $v$ is  ultra-relativistic, we only plotted 
the even-parity PhB structure. Note that  for the perfect
conductor case of panel (e), the ODOS does not present any peaks except for Wood's anomaly and  the PhB structure is completely absent.

Clearly, the calculated uSPR intensity shown in panels (a) and (b) is well correlated with the PhB structure.  Namely, the bright curves of strong SPR 
intensity are   type (B) signals having a positive slope and tracing  very well the even-parity PhB dispersion curves. In addition, we can recognize type (E) signals of  the bright flat lines terminated at the backward light
line, which are seen just   at the frequencies of  the pseudo-gaps.  These features agree  with what we have seen in Sec. III.  Most importantly, the PhB structure with larger dielectric constant is  indeed probed by the uSPR.

As for the Drude  case shown in panels (c) and (d), 
the PhB structure
is composed of  many flat bands.   
These PhB modes  have their origin in  the tight-binding
coupling among cylinders of the surface plasmon polaritons (SPP),  localized on each   cylinder surface \cite{Yannopapas:M:S::60:p5359-5365:1999,vanderLem:M::2:p395-399:2000,Ito:S::6404:p045117:2001}. 
 The calculated intensity map demonstrates that these SPP
bands are  coupled only weakly to the incoming evanescent wave. In contrast, the PhB  
around $\omega a/2\pi c=0.5$, which  has a modest group
velocity, is strongly coupled to the evanescent wave, yielding a very
strong SPR signal.  We thus conclude that  uSPR carries information of PhBs
of SPP origin.

Finally, in Fig. \ref{spr_eps}(e) 
the strong intensity of the SPR arises solely on the shifted $v$ lines.
This reflects the absence of the PhB  structure in the array  of perfect-conductor cylinders.
Thus, we can 
conclude that  the uSPR is peculiar to dielectric and 
metallic PhCs with finite dielectric function  and is
completely absent in the systems without  PhBs.

\subsection{Number of cylinders}

\begin{figure}
\begin{center}
\includegraphics*[width=0.6\textwidth]{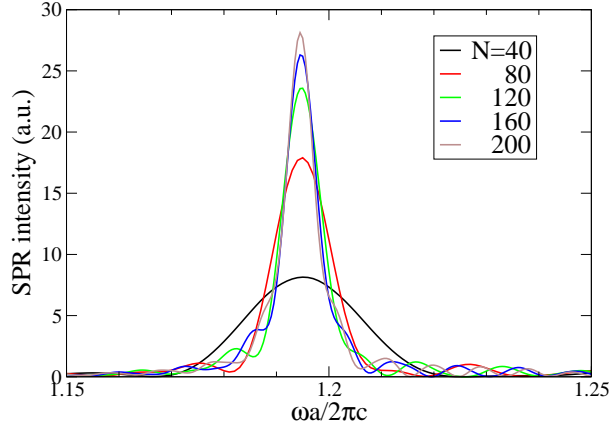}
\end{center}
\caption{\label{N-dependence} The reflected SPR intensity spectra at 
$(\theta,\phi)=(60^\circ,180^\circ)$
 from finite-size PhCs of various $N$. 
The same parameters as in Fig. \ref{conv_teflon} were used except for $N$. }
\end{figure}

In the numerical results shown so far, the number of the cylinders is 
fixed at either $N=21$ or
$N=\infty$.  At small $N$, typically less than 8, the PhB structure
 is not clearly visible in the intensity map of the SPR on the
$(k_x,\omega)$ plane. On the other hand, at large $N$ the PhB structure 
is clearly visible as demonstrated in Figs. \ref{unconv_teflon} and \ref{unconv_teflon_PB}.  
In this region, however,  change of the
SPR intensity map with increasing $N$ is less remarkable. Nevertheless, 
if we have a close look at the spectral line shapes of the uSPR, they 
 indeed change with $N$. 
To investigate this feature, we consider the SPR spectra at 
a fixed solid angle  $(\theta,\phi)=(60^\circ,180^\circ)$ as a function of frequency. 
Figure \ref{N-dependence} shows the spectra for various $N$.  
The SPR signals are strongly enhanced around $\omega a/2\pi c=1.195$, which 
corresponds to an intersection point between the line of 
$k_x-h=(\omega/c)\cos\theta$ (see Eq. (\ref{polar-angle})) 
and the PhB structure
$\omega=\omega_n(k_x,0)$. The intersection point is off the shifted $v$
lines and thus is indeed an uSPR signal. 
We can clearly observe that as $N$ increases, the intensity at the peak 
grows but seems to be saturated to a certain value. 
This implies that the radiation intensity of the uSPR per unit length of the 
electron trajectory decreases at large $N$ and eventually vanishes 
at $N=\infty$. 
This property is reasonable because the uSPR is completely forbidden 
 in the system of perfect translational invariance with $N=\infty$.   
On the other hand, we also found that the intensity of the cSPR on the
shifted $v$ lines increases almost linearly with $N$, as expected from
the conventional theory of the SPR.
Therefore, at very large $N$ the cSPR signals will dominate over 
the uSPR ones. 
However, even at $N=200$ we found that the SPR intensity map does not 
differ so much from Fig. \ref{unconv_teflon}, in which the uSPR signals are rather
stronger than the cSPR ones.      
Besides, in Fig. \ref{N-dependence} we can clearly observe that the
spectral width of the peak decreases with increasing $N$. 
This property reflects better confinement of the radiation energy 
for larger $N$. This width should converge to a certain value at 
$N=\infty$, which is inversely proportional to the
life-time of the relevant photonic band mode.  This is nothing but 
the homogeneous broadening of the spectral line width of the uSPR.

\subsection{Number of stacking layers}

So far we have confined ourselves to the  monolayer PhC.   Now let us stack the identical monolayers periodically in the   $y$ direction. 
As we increase the number $N_l$ of the stacking layers, the ODOS  
reveals  a progressively finer structure  as a function of frequency. 
Each peak of
 ODOS corresponds to a quasi-guided PhB mode confined in
the stacked layers.   The  typical peak-to-peak  distance in frequency 
 is  inversely proportional to $N_l$.    Moreover,  each 
peak is generally getting sharper.

If $N_l$ is large enough, the scattering of the evanescent wave in the ultra-relativistic regime 
 is identical  to the transmission and reflection of  a TE-polarized plane wave that  enters the PhC with its left edge as an entrance surface.  The slab PhC in question  has a finite thickness   $Na$  in the
$x$ direction and has a large  extension  in the $y$ direction with  the entrance surface  parallel to the $yz$ plane. 
The wave vector component $k_\|$ parallel to the
entrance surface  is conserved, and the incident plane wave excites the bulk 
PhB
modes having the same $k_\|$. There is no momentum conservation in the $x$ component; in principle the light excites  any  PhB modes of arbitrary $k_x$.   The scattered wave is decomposed into 
diffraction channels \cite{Sakoda-PC-book}.

Let us examine  the stacked monolayers with   a square lattice of the
cylinders. 
For  uSPR,  $(\pm\Gamma,k_z)$ plays the  role of  $k_\|$
in the above analogy,  and thus we may set  $k_\| \simeq 0$ in the ultra-relativistic case,  provided $k_z=0$.  
Accordingly, the incident plane wave has the wave vectors $(k_x,0,0)$ and propagates in 
the $\Gamma-X$ direction of the square
lattice.    The bulk PhB modes along $\Gamma-X$ 
are thus excited. The conclusion is thus  that  uSPR will carry information   along $\Gamma-X$ of the bulk PhB modes if both $N$ and $N_l$ are sufficiently large.     According to the above
arguments, the SPR intensity  is expected to be enhanced 
in  the forward and backward directions, which correspond to the specular
transmission and reflection.  In addition, if $\omega a/2\pi
c >1$, the SPR intensity is expected to be 
enhanced also on the curves  
\begin{eqnarray}
k_x=\pm\sqrt{\left({\omega\over c}\right)^2-h^2 },  
\label{parabola}
\end{eqnarray}
which correspond to the diffraction channels  associated with the
reciprocal lattice points $h=2\pi n/a \; (n:{\rm integer})$  
along the $k_y$ axis.

\begin{figure*}
\begin{center}
\includegraphics*[width=0.9\textwidth]{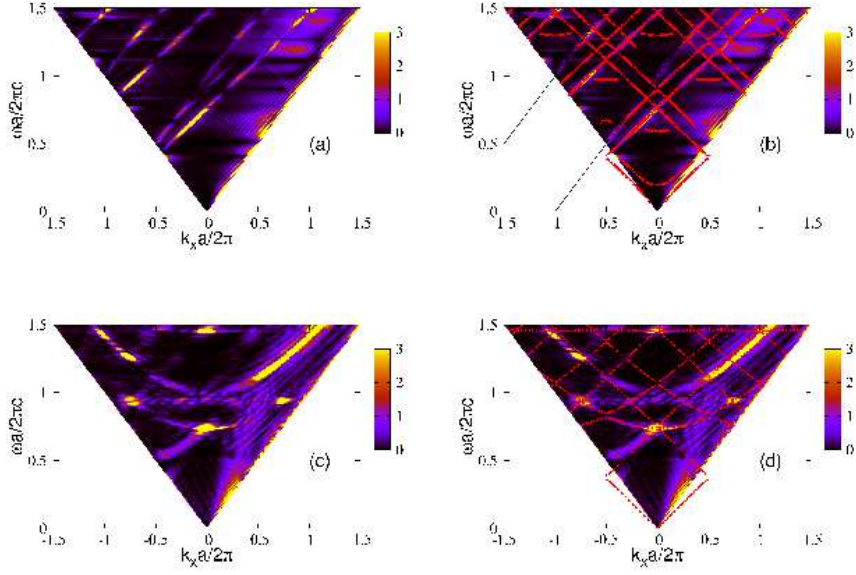}
\end{center}         
\caption{\label{stack} 
 Reflected SPR intensity map of various PhCs of cylinders.
Panel (a) shows the result of  the double-layer system  of contact 
 low-index cylinders ($N=21$ and $N_l=2$). 
Panel (b) is produced from panel (a) by overlaying   the even-parity  PhBs. The mirror plane of the parity 
lies  in between the double-layers.    
Panel (c) shows the result for a  multi-layer PhC  of square lattice of contact 
 low-index cylinders ($N=8$  and $N_l=20$).   Panel (d) is 
a reproduction of panel (c), overlaid  with the PhB  structure along the 
$\Gamma-X$ direction.   Only the even-parity modes with respect to the
 mirror plane relevant to  $\Gamma-X$ are shown. 
The same parameters as in Fig. \ref{conv_teflon} 
were used for the  electron beam. 
} 
\end{figure*}

Figure \ref{stack} illustrates the reflected SPR intensity
maps and the relevant PhB structure of   low-index cylinders in contact.  We used $\varepsilon=2.05$, $v=0.99999c$ and $b=3.33a$.
In Fig. \ref{stack}(a)  the spectrum from  the 
double-layer ($N_l=2$) structure with $N=21$ is shown.  
The intensity map overlaid with  the PhB  structure of 
the double layers (but for $N=\infty$) is shown in 
panel (b).
As before, we plotted only  the even-parity PhB structure.  In the double layer, the mirror plane lies midway  between the layers.  We see the number of bands is  almost
twice that of the monolayer  band structure  shown in Fig. \ref{teflon_band}. This is reasonable, since  the   degenerate   band-structures of each of the monolayers are  split in the double layer.
Obviously there is  a very good correlation of the strong signals of uSPR 
 with the  band structure 
of the even parity.

 Figure \ref{stack}(c) shows the reflected  intensity 
map of the finite multilayers PhC of  $N_l=20$ and $N=8$. 
 We consider this to be  a test  system simulating the slab-type PhC of square lattices. 
We observe at once a  signal of high intensity  along a hyperbolic curve 
 whose bottom is found at $(k_xa/2\pi,\omega a/2\pi c)=(0,1)$. Obviously, this curve
 corresponds to  Eq. (\ref{parabola}) with $h=1$. 
Strong SPR signals other than the hyperbolic curve are found at 
$\omega a/2\pi c=0.73$, 0.93, and 1.46. To identify these signals, 
Fig. \ref{stack}(c) was overlaid with the even-parity PhB structure along 
the $\Gamma-X$ direction of the square lattice. The result is shown 
in Fig. \ref{stack}(d).  
As can be clearly seen, the strong signals correspond to the anti-crossing 
 points of the even-parity PhB structure. 
The bright curve connected to the strong signal around $\omega a/2\pi
c=0.73$ is shown to be along the PhB dispersion curve.
Thus, we can conclude that  
the intensity map of the uSPR 
correlates well with the corresponding PhB structure even in the case 
of stacked monolayers.

\subsection{Azimuthal angle}

So far, we have considered 
the case of $k_z=0$ ($\phi=0^\circ$ and $180^\circ$), that is, we have examined
the radiation emitted within the
$xy$ plane. We here investigate the $\phi$ dependence.  For this purpose,  we  write   the
differential cross section of  SPR in  polar coordinates \cite{Ochiai:O::69:p125106:2004} 
\begin{eqnarray}
& &{\partial W\over \partial\omega\partial\Omega}=
{q\sqrt{q^2-k_z^2}\over 4\pi\mu_0\omega}(|f^M(\theta')|^2+|f^N(\theta')|^2), \label{dcs}\\
& &k_z=q\sin\theta\sin\phi,\quad q={\omega\over c},\\
& &\theta'=-i\log\left( {\cos\theta+i\sin\theta\cos\phi \over 
\sqrt{1-\sin^2\theta\sin^2\phi}}\right).
\end{eqnarray}
Obviously, the cross section must have  inversion symmetry under the operation
$\phi\to -\phi$, reflecting the inversion symmetry of the 
$z$ coordinate with respect to the electron trajectory located at $z=z_0$.

The radiation emission of non-vanishing $k_z$ is generally  small 
compared with that of $k_z=0$,  and  
 $k_z$ is a 
conserved quantity in the scattering by the PhC. Therefore, the $k_z$ dependence of the observed SPR will be controlled dominantly by that of  the decaying 
exponential $\exp(-|\Gamma| b)$ of the initial light. 
This exponential decreases  with increasing $|k_z|$, so that 
the radiation is dominated  by the SPR of   $k_z=0$.

Let us consider the radiation emission toward  solid angle $(\theta,\phi)$. 
In the ultra-relativistic regime it follows  that   
\begin{eqnarray}
\Gamma=\sqrt{\left(\frac{\omega}{c}\right)^2-\left({\omega\over v}\right)^2-(\frac{\omega}{c}\sin\theta\sin\phi)^2}
\simeq iq\sin\theta|\sin\phi|.
\end{eqnarray}
 Thus, the SPR cross section 
at a given $\omega$ and
$\phi(\ne 0^\circ,180^\circ)$ is dominated in the forward ($\theta=0^\circ$) 
and backward ($\theta=180^\circ$) directions.  Similarly, at a given 
$\theta$, the SPR cross section is dominated around  the  
plane perpendicular ($\phi=0^\circ$ and $180^\circ$) to the cylindrical axis.

\begin{figure*}
\begin{center}
\includegraphics*[width=0.9\textwidth]{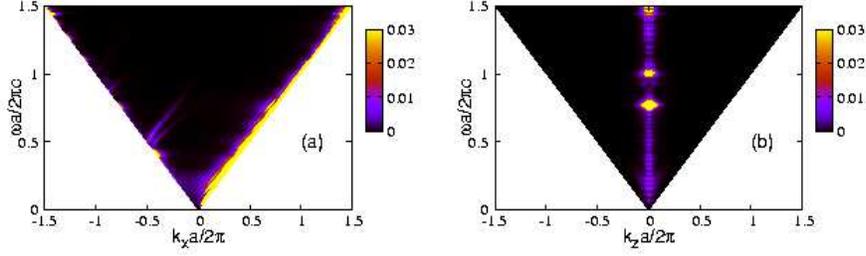}
\end{center}        
\caption{\label{tilte} 
 Reflected SPR intensity off the $xy$ plane.   Spectra from  the monolayer of contact  low-index cylinders are shown.  (a) 
 Intensity  in the $(k_x,\omega)$ plane
 at $\phi=15^\circ$.    (b)  Intensity   in  the $(k_z,\omega)$ plane 
at $\theta=90^\circ$. For the electron beam, the same parameters are
 used  as in Fig. \ref{conv_teflon}.  
} 
\end{figure*}

 Figure \ref{tilte}(a) projects the reflected SPR spectrum
 Eq. (\ref{dcs}) for 
 $\phi=165^\circ$ onto the $(k_x,\omega)$ plane, 
and    Fig. \ref{tilte} (b)  projects the reflected SPR spectrum 
for  $\theta=90^\circ$ is projected onto the $(k_z,\omega)$ plane.  
Figure \ref{tilte}(a) verifies that  the strong intensity 
is limited around the forward and
backward light lines, as asserted above. 
The intensity contrast along the backward light
line is related to the PhB structure with finite $k_z$. 
In  Fig. \ref{tilte}(b), the strong SPR intensity is limited around
$k_z=0$. At $k_z=0$,  three marked peaks can be found at $\omega=0.75,1,$
and 1.5. They correspond to the crossing points between the bright 
curves of Fig. \ref{unconv_teflon}  
and the line of $k_x=0$ (i.e., $\theta=90^\circ$).
Thus, we can conclude that  SPR is highly directive within the plane
normal to the cylindrical axis.

\section{Summary and discussions}
   
To summarize, we have presented a theory of uSPR 
that arises when an ultra-relativistic electron beam is used to obtain the SPR
from a finite PhC composed of cylinders. The ultra-relativistic electron
accompanies an evanescent wave whose spatial decay is almost
negligible at $k_z=0$, so that the evanescent wave can be regarded as a plane wave 
propagating in the  direction of the trajectory. This yields a
peculiar radiation emission from the PhCs, which cannot be explained by
the conventional theory of the SPR in which the finiteness of PhC is treated as infinite.
The spectrum of the uSPR can be used as a probe of the
 PhB structure of the quasi-guided modes having the even-parity symmetry
with respect to the relevant mirror plane.

We have also presented  properties of the uSPR in detail 
by changing several   system parameters. We found that 
the uSPR coexists with the cSPR at moderate
velocities typically in between $0.7c$ and $0.99c$. 
We also found that the uSPR  is completely
absent in the perfect-conductor cylinders because of the absence of 
PhBs. Otherwise, the spectra of the uSPR
correlate with the corresponding PhB structure very well. 
We also found that the cross section of the 
SPR at an ultra-relativistic velocity 
is highly directive within the plane normal to the cylindrical
axis.

It should be emphasized again that the uSPR is 
an unexpected phenomenon, in which the conventional theory assuming
infinite periodicity of the PhCs fails to reproduce its features. 
On the other hand, the
present theory explains very well both the c and u SPRs in a unified manner.   
There are three key items in the uSPR: presence of PhBs, broken
translational invariance of PhC, and ultra-relativistic velocity of
the electron beam. Lack of either one of the three items prevents 
understanding of the uSPR correctly.

In actual PhCs various types of disorder or randomness 
are inevitable, yielding the inhomogeneous broadening of 
spectral line width of the SPR.  For instance, a rigid
vibration of constituent cylinders of the PhC gives rise to 
the Brillouin scattering. As a result, the broadening of the line width  
is given by  the frequency of the vibration.  
The relative percentages of various disorder factors  
depend crucially on the frequency range concerned. 
Thus, when we extract the intrinsic SPR signals from PhC, 
we should carefully take account of disorder.

From the point of view of the radiation emission from high-energy electron  
interacting with periodic structure, we should comment on the peculiarity 
of the uSPR in comparison to the channeling radiation, 
or in other words, Kumakhov radiation  \cite{Kumakhov-book}. The latter radiation occurs inside 
a crystal when a certain condition is satisfied for the incident angle of 
the electron with respect to a major crystal direction. 
The radiation depends strongly on the 
meandering trajectory of the electron trapped around a crystal plane 
or a crystal axis and has a monotonic frequency.  
On the other hand, the uSPR does not require the meandering of the electron 
trajectory. Actually, in our theoretical approach, the trajectory of the electron is assumed to be straight. In addition, the radiation spectrum of the uSPR 
is not monotonic for a fixed trajectory and the typical frequency range is 
inversely proportional to the lattice constant of the PhC under consideration. 
Therefore, the uSPR is not categorized into the channeling radiation.

The uSPR is, in some sense, similar to the transition radiation 
\cite{Landau-EDCM-book} because the broken
translational invariance along the electron trajectory is crucial 
 in both the radiations. However, there is a marked difference in the
directivity between the transition radiation and the uSPR. 
Suppose that an ultra-relativistic
 electron passes from vacuum to a dielectric medium. It is better to 
 focus on the radiation into the vacuum side, because the induced
 radiation in the medium is a mixture of the transition radiation and the
 Cherenkov radiation.  
It was shown that this radiation into the vacuum side is backward-oriented.
On the other hand, as we showed in  the paper, such a high
 directivity into the backward direction is only possible if  the
 relevant frequency lies in a pseudo gap of the PhB structure.   
Among other electron-induced radiations, the uSPR may have  the
 closest resemblance to the diffraction radiation regarding the 
broken translational invariance and  the trajectory which does not pass
 through any air/dielectric interfaces. 
To further clarify the resemblance, a detailed investigation of the
 diffraction radiation in PhC is in order.

\section*{Acknowledgments}  
The authors would like to thank N. Horiuchi, J. Inoue, Y. Segawa, Y. Shibata, 
K. Ishi, Y. Kondo, H. Miyazaki, and S. Yamaguti
for valuable discussions. 
This work was supported by Grant-in-Aid (No. 18656022 for T. O. and 
No. 17540290 for K. O.) 
for Scientific Research from the Ministry of Education, Culture, Sports, Science and Technology.

\end{document}